\documentclass{article}
\usepackage{spconf,amsmath,graphicx,hyperref}
\usepackage{multirow}
\usepackage{booktabs}
\usepackage{xcolor}
\definecolor{Yellow}{rgb}{1, 1, 0.7}
\title{Distillation-based Layer Dropping (DLD): Effective End-to-end Framework for Dynamic Speech Networks}

\name{Abdul Hannan$^{1, 2}$, Daniele Falavigna$^2$, Shah Nawaz$^3$, Mubashir Noman$^4$, Markus Schedl$^{3,5}$, Alessio Brutti$^2$}%\thanks{Thanks to XYZ agency for funding.}}

\address{$^1$University of Trento, Italy, $^2$Fondazione Bruno Kessler, Italy, \\ $^3$Johannes Kepler University Linz, Austria, $^4$MBZUAI U.A.E., \\ $^5$Linz Institute of Technology, Austria}

\begin{document}
%\ninept
%
\maketitle
\begin{abstract}
% Dynamic architectures, because of architectural adaptability, thrive in constrained and varying resource settings, which is common for edge and IoT devices.
% Dynamic architectures succeed in constrained and varying resource settings, which is a common occurrence for edge and IoT devices, because of their architectural adaptability. Layer Dropping, which skips parts of the network to reduce computational complexity,  

Edge devices operate in constrained and varying resource settings, requiring dynamic architectures that can adapt to 
%fluctuations in 
limitations of the available resources. 
To meet such demands, layer dropping ($\mathcal{LD}$) approach is typically used to transform static models into dynamic ones by skipping parts of the network along with reducing overall computational complexity. However, existing $\mathcal{LD}$ methods greatly impact the dynamic model's performance for low and high dropping cases, deteriorating the performance-computation trade-off. %struggling to exploit the computational capacity of full model. %with acceptable performance in middle range.
%To this regard, we propose an enhanced framework using knowledge distillation in parallel with $\mathcal{LD}$ that supervises the output of dynamic model to overcome the performance limitations of the state-of-the-art methods. 
To this end, we propose a distillation-based layer dropping (DLD) framework that effectively combines the capabilities of knowledge distillation and $\mathcal{LD}$ in an end-to-end fashion, thereby achieving state-of-the-art performance for dynamic speech networks. Comprehensive experimentation utilizing well-known speech recognition methods, including conformer and WavLM, on three public benchmarks demonstrates the effectiveness of our framework, reducing the word error rate by $9.32\%$ and $2.25\%$ for high and no dropping cases with $33.3\%$ reduction in training time.

%that benefits from the rich semantic output of reference model to overcome the performance degradation in state-of-the-art. out-performs the
% The abstract should appear at the top of the left-hand column of text, about 0.5 inch (12 mm) below the title area and no more than 3.125 inches (80 mm) in length.  Leave a 0.5 inch (12 mm) space between the end of the abstract and the beginning of the main text.  The abstract should contain about 100 to 150 words, and should be identical to the abstract text submitted electronically along with the paper cover sheet.  All manuscripts must be in English, printed in black ink.
\end{abstract}

\section{Introduction}
\label{sec:intro}
Speech foundation models offer promising performance on various downstream tasks such as automatic speech recognition (ASR), spoken language understanding, speaker identification etc., due to their rich semantic representation capability. 
%This efficient performance comes at the cost of enormous computational complexity limiting their deployment on limited and varying resource devices. 
This favorable performance is achieved at the cost of enormous computational complexity limiting their deployment on the resource-constrained devices with resource variability.
Conventional techniques like pruning \cite{lecun1989optimal, zhao2021progressive}, quantization \cite{jacob2018quantization} and knowledge distillation \cite{hinton2015distilling} renders compression with relatively good performance-computation trade-off, however, these methods do not suffice the requirements of a dynamic architecture that is preferred for varying-resource devices. To achieve dynamic architectures that are well-suited for such scenarios, techniques like \textit{data-offloading:} presents privacy and latency issues \cite{kumar2013survey, matsubara2022split}, \textit{early exit:} involves delicate design choices for inserting exit branches \cite{branchynet2017}, and especially \textit{layer dropping (or skipping)} are utilized enabling the architecture to perform efficiently in such dynamic environment.%produce output using variable model sizes. 

Layer dropping $\mathcal{(LD)}$, inspired from stochastic depth \cite{huang2016deep} in vision domain, has been increasingly utilized to transform static architectures into dynamic ones \cite{huang2016deep, fan2019reducing, Wu2018BlockDrop}. 
%$\mathcal{LD}$, also referred as structured pruning, executes some parts of the network and skips the rest, providing regularization during training phase and achieving faster execution during training and inference phase. $\mathcal{LD}$ can be categorized by the strategy employed to drop subnets: (i) \textit{random dropping} (RD): where each module can be dropped depending on probability \cite{fan2019reducing, zhang2020accelerating, zaiem2023fine, hannan2024}, and (ii): \textit{data-driven}: where input of the model constitutes the dropping blueprint \cite{Wu2018BlockDrop, chen2019you, peng2023i3d, hannan2025input, genova2025keep}.
$\mathcal{LD}$, also referred as structured pruning, executes some parts of the network and skips the rest, providing regularization during training phase and improving execution time, and can be categorized as: (i) \textit{random dropping} (RD): where each module can be dropped depending on probability \cite{fan2019reducing, zhang2020accelerating, zaiem2023fine, hannan2024}, and (ii): \textit{data-driven}: where input of the model constitutes the dropping blueprint \cite{Wu2018BlockDrop, chen2019you, peng2023i3d, hannan2025input, genova2025keep}. Zhang and He \cite{zhang2020accelerating} proposed a gating scheme for RD to skip different subnets of pre-trained networks using BERT, RoBERTa, and XLNet architectures. In context of ASR domain, recent work \cite{hannan25b_interspeech} presented an effective two-step framework to create multiple small sized static models, however, the training recipe requires enhancement as well as the resulting models are not suitable for varying resource environments. Another study \cite{zaiem2023fine} compared the performance of different approaches, i.e., early exit, input downsampling, and RD with a dropping probability $p_d = 0.5$, by fine-tuning a pre-trained network on a subset of dataset. However, they omit the evaluation for extreme dropping scenarios. LDASR \cite{hannan2024} investigated the inference time behavior of conformer architecture trained using RD with $p_d = \{0.2, 0.5, 0.8\}$, and revealed that $p_d=0.5$ provides optimum performance-computation trade-off. However, their approach offers substantial performance degradation for high dropping values as well as struggles to exploit the computational capacity of the complete model in case of no (or low) dropping. Similarly, other works attempts to achieve a dynamic network using learnable masking \cite{genova2025keep}, and data-driven methods \cite{peng2023i3d, xu24_interspeech}, and either omit evaluation or offer significant performance drop for high dropping values.

%############### Figure ####################
\begin{figure*}[!t]
    \centering
    \includegraphics[width=0.8\linewidth]{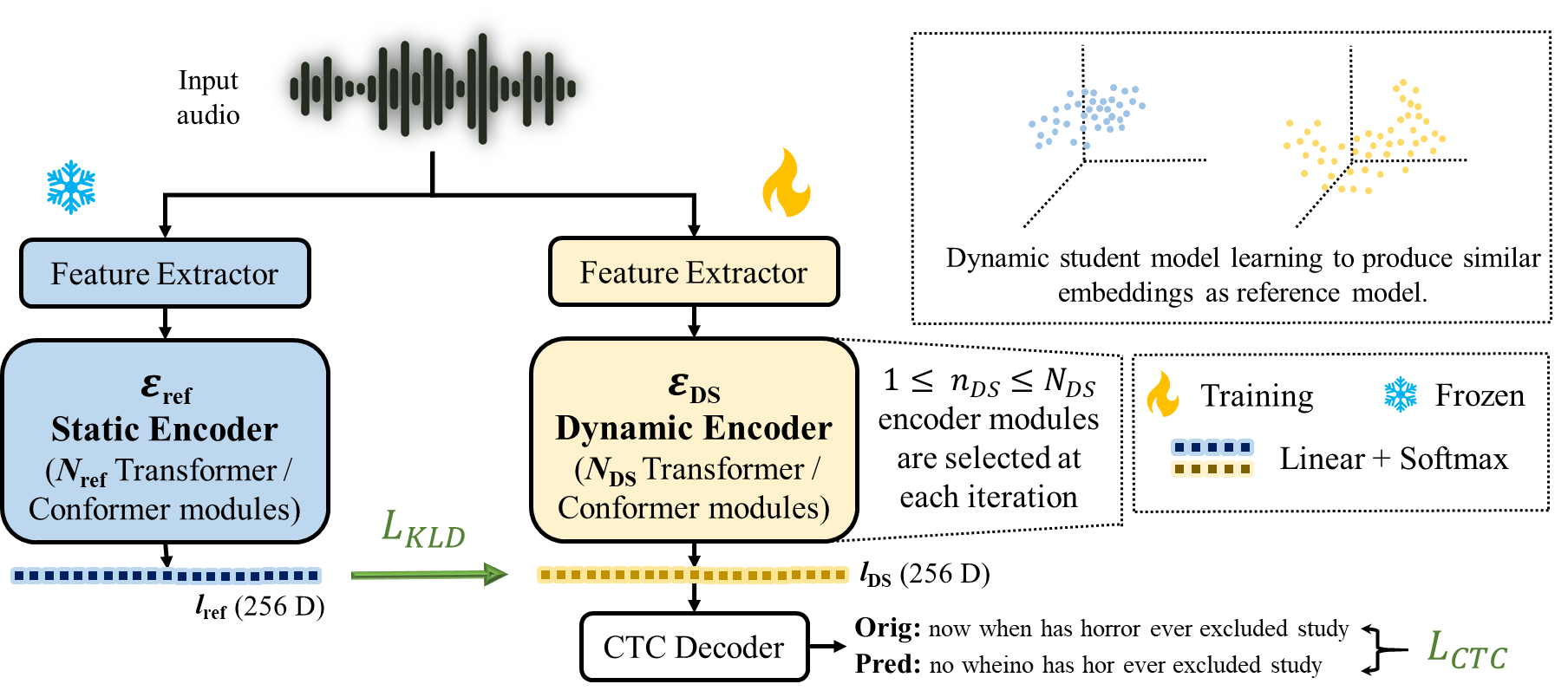} \vspace{-1em}
    \caption{Illustration of proposed DLD framework. $\mathcal{M}_\text{ref}$ uses all encoder layers to supervise the embeddings of $\mathcal{M}_\text{DS}$.} %\vspace{-2mm}
    \label{fig:comparison}
    \vspace{-1em}
\end{figure*}

Besides, knowledge distillation (KD) has been extensively utilized to achieve model compression for ASR downstream task, resulting in compressed static models. TutorNet \cite{yoon2021tutornet} performs KD from multiple locations of teacher model, whereas others \cite{chebotar2016distilling, yang2023knowledge} utilized several teacher networks to supervise the student model. 
%To the best of our knowledge, this is the first study to employ KD with dynamic architectures for ASR. 
In this regard, inspired from two-step framework of \cite{hannan25b_interspeech}, this work aims to address the aforementioned issues of dynamic architectures using knowledge distillation in conjunction with random dropping and is the first of its kind to employ KD with dynamic architectures for ASR to the best of our knowledge.

To summarize, our contributions are: \vspace{-1mm} 
\begin{enumerate}
    \item We propose an end-to-end DLD framework that distills expert knowledge of large model to the dynamic student model, featuring promising performance. \vspace{-1.5mm}
    \item The introduced approach resolves the performance degradation issue for no dropping, in addition to improving overall performance for all model sizes. \vspace{-1.5mm}
    \item We perform comprehensive experimentation on two publicly available ASR benchmarks using conformer and WavLM architectures, demonstrating the superiority of proposed framework. \vspace{-1.5mm}
\end{enumerate}
% We propose an end-to-end framework that distills expert knowledge of static model (called as reference model) to direct the output of a dynamic student model produced using varying encoder modules. As a result, the student model yields improved final embeddings resulting in better transcriptions, providing better performance-computation trade-off.
%strives to generate improved embeddings 

%Although, RD transforms static model into dynamic ones, the performance degradation for high dropping values is substantial and requires improved approaches to mitigate it. Secondly, dynamic architectures fail to leverage from the rich semantic representation of full model in case of no (or low) dropping. 
% \textcolor{red}{First Paragraph -- Large models deployment on low-resource devices,,,, Static models to Dynamic model conversion and its issues...... \\
% Layer Dropping provides dynamic model, but comes with a performance gap when no blocks are dropped. \\
% Training the model in end-to-end fashion along with aligning the distribution of embeddings using knowledge distillation. }
% \section{Background}
% \label{sota}

\section{Methodology}
\label{method}
To enhance the inference time efficacy of a dynamic model and mitigate performance degradation issues, we propose an end-to-end DLD\footnote[1]{Our code is available online \textcolor{blue}{https://github.com/hannabdul/DLD4ASR}} framework that aligns the distribution of latent embeddings of a dynamic student model $\mathcal{M}_\text{DS}$ to a reference model $\mathcal{M}_\text{ref}$. 
%The alignment performed minimizes the Kullback-Leibler divergence (KLD) loss between LSRs of 
 
\vspace{1mm} \noindent \textbf{Architecture and Data Flow:} $\mathcal{M}_\text{ref}$ consists of a feature extractor $\mathcal{F}_\text{ref}$, encoder $\mathcal{E}$ with $N_\text{ref}$ transformer/conformer modules, and a linear projector whose output undergoes a softmax operation to get latent embedding $l_\text{ref}$. Similarly, $\mathcal{M}_\text{DS}$ comprises of a feature extractor $\mathcal{F}_\text{DS}$, encoder $\mathcal{E}_\text{DS}$ with a total of $N_\text{DS}$ transformer/conformer modules from which $n_\text{DS}$ are used per iteration, and a linear projector followed by a softmax, providing latent embedding $l_\text{DS}$. The $\mathcal{E}_\text{DS}$ is equipped with a gated mechanism that allows to process or skip the $i$-th encoder module using a gate $g^i$. The gate $g^i \in \{ 0, 1 \} $ follows the bernoulli distribution (BD) with a probability of 0.5.
\begin{equation}
    % y^{i+1} = g^i \times y^i 
    y^{i+1} = y^i + g^i \cdot f^i(y^i)
\end{equation}
%where $y^i$ and $y^{i+1}$ are the input and output of the $i$-th encoder module. 
where $y^i$ and $y^{i+1}$ are the input and output of $i$-th encoder block, and $f^i(.)$ is the transformation applied by the $i$-th encoder. As $g^i$ follows the bernoulli distribution, $n_\text{DS}$ also becomes a bernoulli random variable whose value can be determined as $n_\text{DS} = \sum_{i=1}^{N_\text{DS}} g^i$ for each iteration during training period. At inference, the value of $n_\text{DS}$ is defined enabling us to evaluate the model's performance to different encoder depths, i.e. $n_\text{DS} = 2, 4,  \dots ,N_\text{DS}$.

For a dataset $\mathcal{D} \in \{ (a^j, t^j) \}_{j=1}^J$ comprising of $J$ audio-transcription pairs, $\mathcal{M}_\text{ref}$ produces a latent space representation $l_{ref}^j | \mathcal{E}_{N_\text{ref}}$ for the $j$-th input using $N_\text{ref}$ encoder modules. % that serves as the target representation for the dynamic student model. 
In contrast, $\mathcal{M}_\text{DS}$ utilizes $n_\text{DS}$ number of encoder modules for each iteration to produce the latent space representation $l_\text{DS}^j | \mathcal{E}_{n_\text{DS}}$ where $1 \leq n_\text{DS} \leq N_\text{DS}$.

\vspace{1mm} \noindent \textbf{Objective Function:} To align the distributions of reference model's embedding $l_\text{ref}^j|\mathcal{E}_{N_\text{ref}}$ and dynamic student model's embedding $l_\text{DS}^j|\mathcal{E}_{n_\text{DS}}$, we employ Kullback–Leibler divergence ($\mathcal{L}_{\texttt{KLD}}$) loss which minimizes the statistical distance between both embeddings.
%The latent space representations of reference model and dynamic student model are aligned using Kullback–Leibler divergence (KLD) loss. The KLD loss minimizes the difference between distribution of $l_{ref}^j|\mathcal{E}_{N_{ref}}$ and $l_{DS}^j|\mathcal{E}_{n_{DS}}$ embeddings produced using frozen reference model and varying-sized student model.
% The latent space representations of reference model are 
\vspace{-2mm}
\begin{equation}
    \mathcal{L}_{\texttt{KLD}} = \text{min} \sum_{j=i}^{J} \text{D}_{\texttt{KL}} \bigl(  l_\text{ref}^j \: || \: \:  l_\text{DS}^j \bigr)
\end{equation} 
In addition, the token probabilities $l_\text{DS}^j|\mathcal{E}_{n_\text{ DS}}$ produced by $\mathcal{M}_\text{DS}$ for input $a^j$, are forwarded to a Connectionist Temporal Classification (CTC) decoder \cite{graves2006connectionist} to generate predicted text, which is employed to estimate CTC loss ($\mathcal{L}_{\texttt{CTC}}$) by comparing with ground-truth transcription $t^j$.
\vspace{-2mm}
\begin{equation}
    \mathcal{L}_{\texttt{CTC}} = \text{min} \sum_{j=1}^{J} \mathbf{F}_{\mathtt{CTC}} \big( l_\text{DS}^j|\mathcal{E}_{n_\text{ DS}} \:, t^j \big)
\end{equation}
The overall loss function to be minimized is given as:
\begin{equation}
    \mathcal{L} = \mathcal{L}_{\texttt{KLD}} + \mathcal{L}_{\texttt{CTC}}
\end{equation}

%$\mathcal{M}_\text{DS}$ and $\mathcal{M}_\text{ref}$ consist of a feature extractor, $N$ encoder modules, and a linear projection layer where $\mathcal{M}_\text{DS}$ uses $1 \leq n\leq N$

%#########################################
\begin{table}[t]
        \setlength{\tabcolsep}{2pt}
        \caption{Comparing WER (in \%) of proposed framework on conformer architecture against RD based baseline methods when trained on LibriSpeech 1000. RD - random dropping, Params column depicts number of executed parameters.}
        \vspace{1mm}
        \begin{tabular}{cccccccc}
            \toprule
            \multirow{2}{*}{\textbf{$n_\text{DS}$}} & \multicolumn{3}{c}{\textbf{LibriSpeech}} & \multicolumn{3}{c}{\textbf{TEDLIUM v3}} & \textbf{Params} \\ \cmidrule(l){2-4} \cmidrule(l){5-7}
            & \textbf{$\text{RD}_\text{sc}$} & \textbf{$\text{RD}_\text{LD}$} & \textbf{Ours} & \textbf{$\text{RD}_\text{sc}$} & \textbf{$\text{RD}_\text{LD}$} & \textbf{Ours} & \textbf{M} \\ \midrule%\cmidrule(r){1-1}
            12  & 8.07 & 7.69 & \colorbox{Yellow}{5.82} & 13.72 & 14.03 & \colorbox{Yellow}{12.36} & 31.2 \\
            10  & 7.28 & 6.95 & \colorbox{Yellow}{5.90} & 14.01 & 13.94 & \colorbox{Yellow}{12.69} & 26.06 \\
            8  & 7.28 & 6.92 & \colorbox{Yellow}{6.32} & 15.12 & 14.25 & \colorbox{Yellow}{13.05} & 20.91 \\
            6  & 8.39 & 7.98 & \colorbox{Yellow}{7.32} & 17.24 & 15.78 & \colorbox{Yellow}{14.89} & 15.76 \\
            4  & 12.81 & 12.54 & \colorbox{Yellow}{11.81} & 24.46 & 22.43 & \colorbox{Yellow}{20.88} & 10.61 \\
            2 & 39.87 & 43.62 & \colorbox{Yellow}{38.30} & 54.20 & 55.30 & \colorbox{Yellow}{51.40} & 5.47 \\ \bottomrule
            $\mathcal{M}_\text{ref}$ & \multicolumn{3}{c}{5.29} & \multicolumn{3}{c}{11.83} & 31.2\\ \bottomrule
        \end{tabular}
    \label{tab:conf}
    \vspace{-3mm}
\end{table}

\section{Experimentation and Results}
\label{exp}

\noindent \textbf{Architectures:} We evaluated the proposed DLD framework using two architectures: (i) Conformer: a modified version with light feature extractor along with linear projector instead of LSTM-based projector (similar to \cite{hannan2024, hannan2025input}), (ii) WavLM-base model \cite{chen2022wavlm}. For conformer model, mel-spectrograms are extracted for each 320 samples (20 ms) of input sequence with a hop-length of 160 samples (10 ms), whereas the input sequence is fed directly to the WavLM model.

\noindent \textbf{Implementation Details:} For both architectures, we kept the reference model frozen and finetuned the dynamic student model using CTC loss on a NVIDIA $A40$ GPU. In case of Conformer, the weights of $\mathcal{M}_\text{DS}$ are initialized with pre-trained weights of static model and finetuned for $100$ epochs using batch size of $64$ and L2 regularization of $5e^{-4}$. The learning rate is increased for 10k warm-up steps and exponentially decreased till the end of training. In case of WavLM-base, we initialized $\mathcal{M}_\text{DS}$ with pre-trained weights and finetuned using Adam optimizer with a batch size of 8 and default configuration settings. We measured the architecture's performance in terms of Word Error Rate (WER).

%\vspace{1mm} 
\noindent \textbf{Datasets:} For ASR downstream task, we employed two publicly available corporas: LibriSpeech-1000 \cite{panayotov2015librispeech} and TED-LIUM v3 \cite{hernandez2018ted}.

%\vspace{1mm} 
\noindent \textbf{Baselines}: For conformer architecture, we compare against Random Dropping (RD) method with dropping probability $p_{d} = 0.5$ \cite{hannan2024} (referred as ``$\text{RD}_\text{LD}$" in this paper) that is trained for $300+$ epochs and an input-driven dropping strategy \cite{peng2023i3d} (referred as ``I3D") trained on $100$ hours of librispeech corpora. Additionally, we compare against a conformer trained from scratch ``$\text{RD}_\text{sc}$" with $p_d = 0.5$ for 150 epochs. For WavLM, we compared against RD with $p_{d} = 0.5$ \cite{hannan2025input} (referred as ``$\text{RD}_{w-sc}$" in this paper). %% and an efficient subnetwrk extraction method \cite{genova2025keep} that uses wave2vec $2.0$ model.
Unless stated, all training procedures are performed using complete datasets $\big($ librispeech ($1000$hrs) and tedlium-v3 ($452$hrs) $\big)$.

\subsection{Results}
\label{res}
% Table \ref{tab:1} and \ref{tab:2} enlists the evaluated WER using aforementioned architectures on LibriSpeech test-clean and TED-LIUM v3 test splits. It is evident from Table \ref{tab:1} that, using conformer architecture, our framework outperforms the models trained without aligning embeddings, providing improved performance-computation trade-off for all encoder sizes, as well as resolves the full-inference performance degradation issue observed in previous works $\text{RD}_{sc}$ and $\text{RD}_{LD}$ \cite{hannan2024}. For instance, performing inference without dropping any encoder module ($n_{DS}=12$) results in WER of $5.82\%$ that is $2.25\%$ less than baseline
\subsubsection{Conformer}
Table \ref{tab:conf} and \ref{tab:wav} enlists the measured WER using aforementioned architectures on LibriSpeech test-clean and TED-LIUM v3 test sets. For conformer architecture, it is evident from Table \ref{tab:conf} that our framework, leveraging from the knowledge distillation to train the dynamic student model, outperforms models trained without such knowledge transfer. 
%It should be noted that
We notice that $\text{RD}_\text{sc}$ and $\text{RD}_\text{LD}$ models are capable of adapting to  varying resource settings, however, their performance degrades when evaluated with less or no dropping ($n_\text{DS} \geq 8$), %which is resolved by our framework. 
which is superbly resolved by proposed DLD. 
We observe a trend of improved performance-computation trade-off on test splits of both datasets, in addition to retaining performance with low or no inference time dropping ($2.25\%$ and $1.67\%$ less WER for $n_\text{DS} = 12$ without dropping any encoder module from respective baseline models). Moreover, for $n_\text{DS} = 6$, our framework achieves $2.35\%$ and $0.89\%$ improved WER with $2$x computational speed-up on TED-LIUM v3 test split. Similarly, when trained on librispeech-100 split, our method outclass I3D (input-driven strategy) \cite{peng2023i3d} that contains transformer-based encoder structure with deep feature extractor and triple number of encoder layers. They report a WER of $\approx$13.5\% and $\approx$12.2\% for dropping $50\%$ and $25\%$ of encoder modules for librispeech test-clean split, which is $\approx3.56\%$ and $\approx4.06\%$ higher than the proposed framework.

% 4.91 WER for 25\% drop  ---- 5.28 \cite{genova2025keep} \\
% 6.73 WER for 50\%  ---- 11.3 \cite{genova2025keep} \\
% 23.25 WER for 75\%  ---- 33.9 \cite{genova2025keep} \\
%############### Table ####################
\begin{table}[t]
    \setlength{\tabcolsep}{2pt}
    \caption{Comparing measured WER (in \%) for finetuning WavLM with RD with and without DLD framework.}
    \vspace{1.5mm}
    \begin{tabular}{@{}ccccccc@{}}
        \toprule
        \multirow{2}{*}{\textbf{$n_\text{DS}$}} & \multicolumn{2}{c}{\textbf{LibriSpeech}} & \multicolumn{2}{c}{\textbf{TEDLIUM v3}} & \textbf{Params} & \textbf{Speed-up} \\ \cmidrule(l){2-3} \cmidrule(l){4-5}
        & \textbf{$\text{RD}_{w-sc}$} & \textbf{Ours} & \textbf{$\text{RD}_{w-sc}$} & \textbf{Ours} & \textbf{M}\\ \midrule%\cmidrule(r){1-1}
        12  & 5.47 & \colorbox{Yellow}{4.57} & 10.32 & \colorbox{Yellow}{9.19} & 94.40 & 1x \\
        10  & 5.78 & \colorbox{Yellow}{4.66} & 10.75 & \colorbox{Yellow}{9.24} & 80.22 & 1.17x \\
        8  & 6.52 & \colorbox{Yellow}{5.20} & 12.53 & \colorbox{Yellow}{10.55} & 66.04 & 1.43x \\
        6  & 9.01 & \colorbox{Yellow}{6.73} & 17.74 & \colorbox{Yellow}{13.89} & 51.87 & 1.82x \\
        4  & 17.59 & \colorbox{Yellow}{12.76} & 31.90 & \colorbox{Yellow}{24.61} & 37.69 & 2.50x \\
        2 & 60.10 & \colorbox{Yellow}{50.78} & 76.76 & \colorbox{Yellow}{68.79} & 23.51 & 4.01x \\ \bottomrule
        $\mathcal{M}_\text{ref}$ & \multicolumn{2}{c}{3.5} & \multicolumn{2}{c}{10.12} & 94.40 & 1x\\ \bottomrule
    \end{tabular}
    \label{tab:wav}
    \vspace{-1.5mm}
\end{table}

\subsubsection{WavLM}
To transform a static WavLM foundation model into a dynamic one, our framework significantly improves the performance-computation trade-off as illustrated in Table \ref{tab:wav}. For instance, we achieve $4.83\%$ and $7.29\%$ better WER on LibriSpeech and TEDLIUM datasets than the baseline with random dropping. Furthermore, our framework surpasses learnable masking method proposed in \cite{genova2025keep} by $0.37\%$, $4.57\%$, and $10.65\%$ for dropping of $25\%$, $50\%$, and $75\%$ encoder modules. 
We observe from Figure~\ref{fig:comparison} that \cite{genova2025keep} also provides reduction in model size, however, they employed Wave2Vec2 foundation model containing $317 M$ parameters ($3.37$x more than ours) along with a bi-LSTM layer (linear projection in our case). The number of utilized parameters of their $75\%$ trimmed model is similar to our full-sized model with substantially low performance ($33.9\%$ vs $4.57\%$). Figure \ref{fig:comparison} illustrates the variation in computational load and performance-computation trade-off between WavLM (ours) and wav2vec2 \cite{genova2025keep}. 

\begin{table}[t]
    \centering
    \caption{Evaluation of proposed framework on LibriSpeech test-clean split using Conformer model as training progresses.}
    \vspace{1mm}
    \begin{tabular}{ccccccc}
        \toprule
        \multirow{2}{*}{\textbf{$n_\text{DS}$}} & \multirow{2}{*}{$\text{RD}_\text{sc}$} & \multirow{2}{*}{$\text{RD}_\text{LD}$} & \multicolumn{4}{c}{\textbf{WER at Epoch}} \\ \cmidrule{4-7} 
         & & & \textbf{25} & \textbf{50} & \textbf{75} & \textbf{100} \\ \midrule
        12 & 8.07 & 7.69 & 6.39 & 6.09 & 5.89 & 5.82 \\
        10 & 7.28 & 6.95 & 6.47 & 6.19 & 6.04 & 5.90 \\
        8 & 7.28 & 6.92 & 6.93 & 6.64 & 6.40 & 6.32 \\
        6 & 8.39 & 7.98 & 8.16 & 7.74 & 7.49 & 7.32 \\
        4 & 12.81 & 12.54 & 13.26 & 12.58 & 12.09 & 11.81 \\
        2 & 39.87 & 43.62 & 41.26 & 39.77 & 38.82 & 38.30 \\ \bottomrule
    \end{tabular}
    \label{tab:abl-1}
    \vspace{-2mm}
\end{table}

\noindent \textbf{On performance degradation }: Table \ref{tab:conf} highlights that DLD framework resolves the inference time degradation problem present in the conformer baseline methods ($\text{RD}_\text{sc}$ and $\text{RD}_\text{LD}$ \cite{hannan2024}). For $n_\text{DS} = \{12, 10, 8\}$, our framework scores WER of $\{5.82, 5.90, 6.32\}$ with minimal performance drop for no and low dropping values, whereas $\text{RD}_\text{sc}$ and $\text{RD}_\text{LD}$ gives WER of $\{8.07, 7.28, 7.28\}$ and $\{7.69, 6.95, 6.92\}$ respectively, depicting substantial performance degradation. We attribute this performance limitation of baseline methods to the shallow feature extractor which fails to provide rich-semantic representation for the input audio. Yet, our framework produces efficient results with two-third number of epochs of what are required to train the model without our framework from scratch. We also conjecture that increasing the depth of feature extractor can mitigate this issue as shown in the case of WavLM model (see Table \ref{tab:wav}) where our framework maintains its superiority over the baseline $\text{RD}_\text{w-sc}$ \cite{hannan2025input}. 

\subsection{Ablations} \label{ablations}

\subsubsection{Framework performance evaluation as a function of training time}
To establish the efficiency of proposed framework, we compute WER for different encoder sized models over the course of training, i.e. at epoch 25, 50, 75 and 100. Table \ref{tab:abl-1} demonstrates that our framework surpasses the baseline models (referred as $\text{RD}_\text{sc}$ and $\text{RD}_\text{LD}$ in Table \ref{tab:conf}) at epoch $50$ that is $\leq \frac{1}{3}$ of the training time required for the baseline methods. This also illustrates that embedding's alignment during training yields fast convergence and improves overall performance.

%############### Figure ####################
\begin{figure}[t]
    \centering
    \includegraphics[width=0.94\columnwidth]{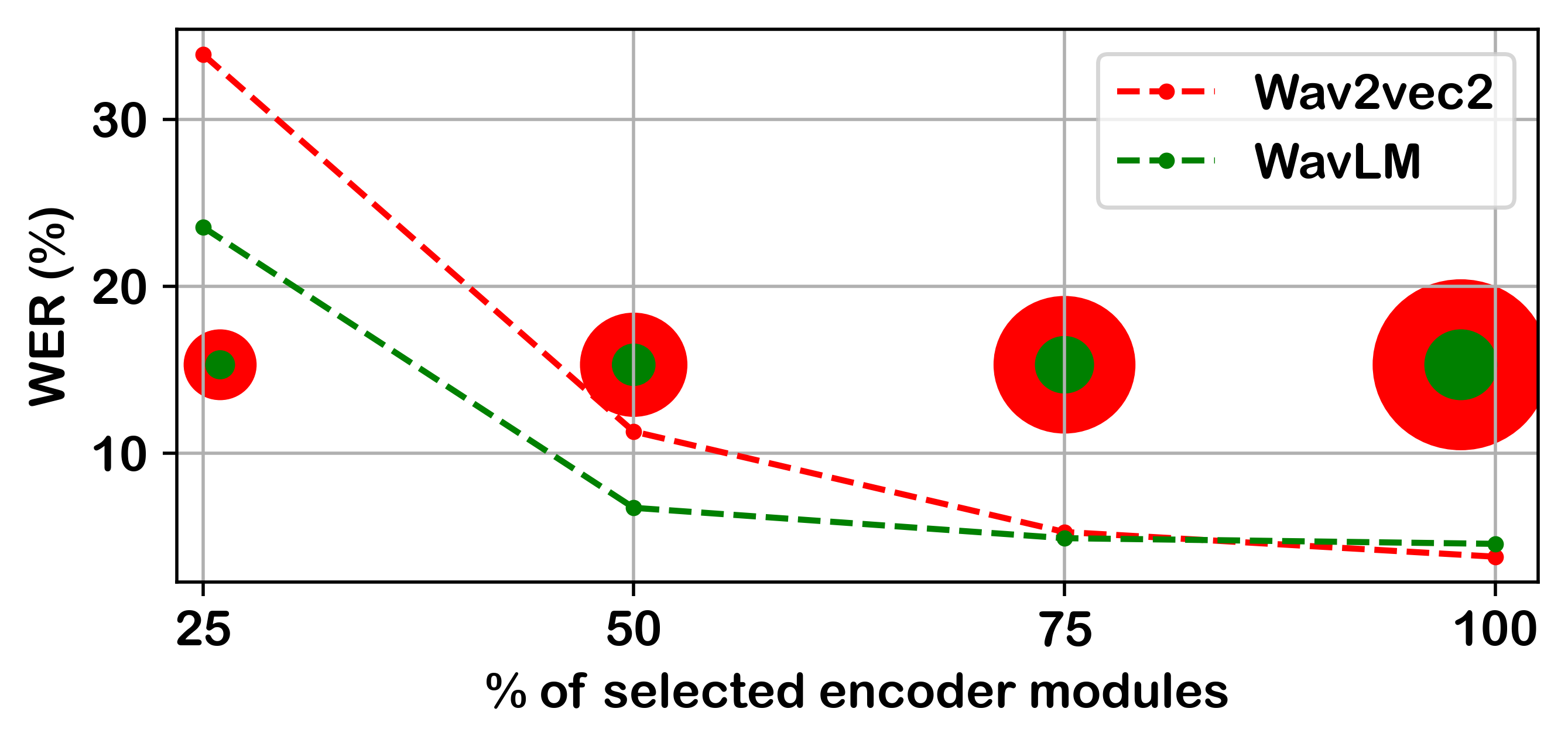} 
    \vspace{-2mm}
    \caption{Comparing dynamic versions WavLM (random dropping based) vs Wav2Vec2 (learnable masking based). Circle depicts the utilized parameters for different encoder sizes.}
    \label{fig:comparison}
    \vspace{-2mm}
\end{figure}

\subsubsection{Generalization of proposed framework on other downstream tasks} 
\begin{figure}[t]
    \centering
    \includegraphics[width=0.95\columnwidth]{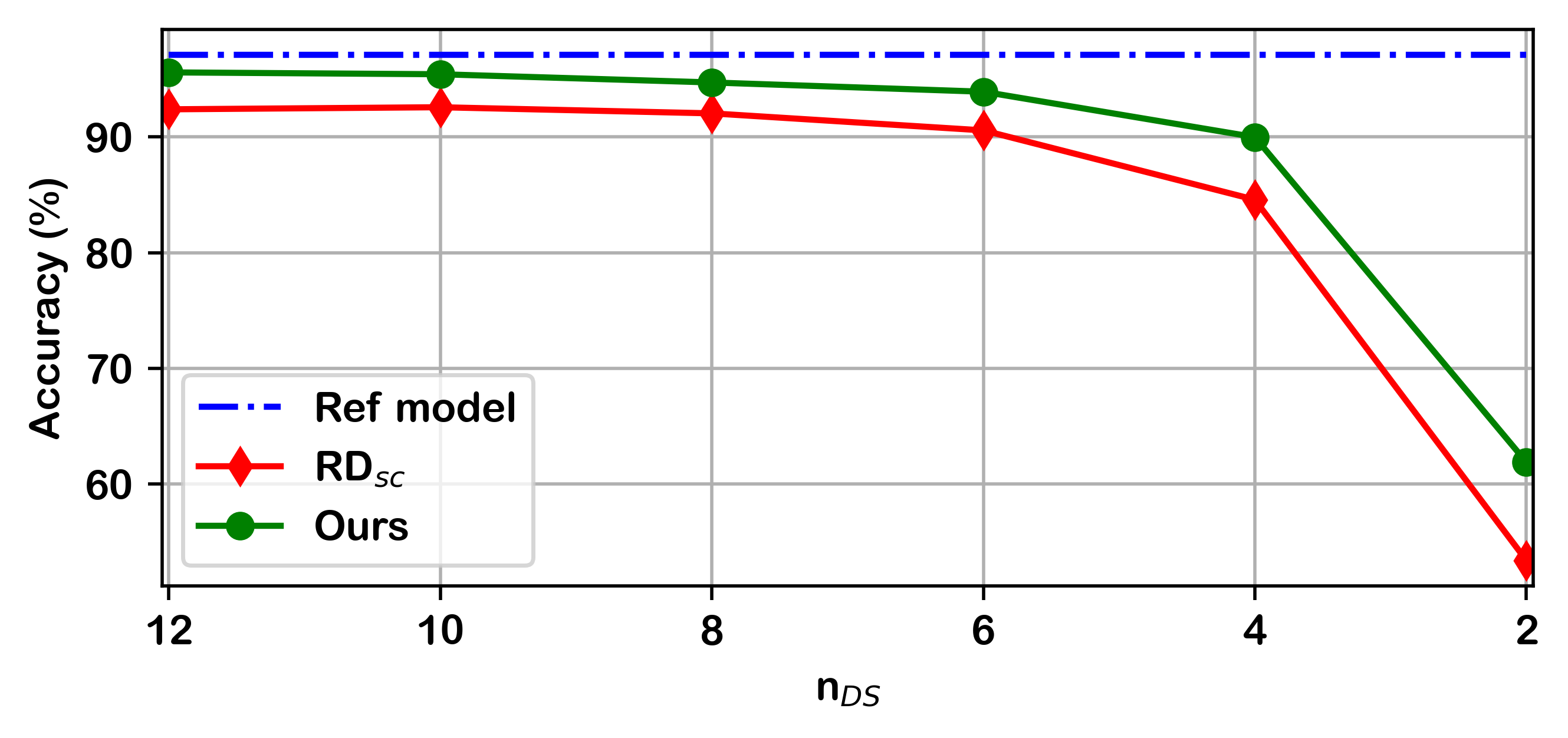} \vspace{-2mm}
    \caption{Evaluating generalization capability of our framework on spoken language understanding task using FSC dataset. Baseline Accuracy (with full model $\mathcal{M}_\text{ref}$ = 97.1\%)}
    \label{fig:abl-2}
    \vspace{-4mm}
\end{figure}

To signify the generalizability of our approach, we adapted the conformer architecture for spoken language understanding task, and evaluated it on fluent speech commands (FSC) \cite{DBLP:conf/interspeech/LugoschRITB19} dataset. For the reference model, we trained the conformer architecture from scratch achieving a baseline accuracy of $97.1\%$ which is in-line with original paper. Similarly, we trained the conformer using RD with $p_d=0.5$ and our KD-based framework. Figure \ref{fig:abl-2} presents the accuracy for different value of $n_\text{DS}$, signifying that effectiveness of DLD is not limited to ASR downstream task only, rather it can be applied to different applications yielding improved performance.
\vspace{-1mm}
%############### Figure ####################

% \begin{table}[t]
%     \centering
%     \begin{tabular}{ccc}
%         \toprule
%         \multirow{2}{*}{\textbf{$n_\text{DS}$}} & \multicolumn{2}{c}{\textbf{RD with $p_d=0.5$}} \\ \cmidrule{2-3} 
%         & \textbf{without KD} & \textbf{\begin{tabular}[c]{@{}l@{}}with KD\\ (Our framework)\end{tabular}} \\ \midrule
%         12 & 92.38 & \colorbox{Yellow}{95.57} \\
%         10 & 92.56 & \colorbox{Yellow}{95.41} \\
%         8 & 92.03 & \colorbox{Yellow}{94.70} \\
%         6 & 90.56 & \colorbox{Yellow}{93.90} \\
%         4 & 84.57 & \colorbox{Yellow}{89.98} \\
%         2 & 53.41 & \colorbox{Yellow}{61.90} \\ \bottomrule
%         \end{tabular}
%     \caption{Evaluating generalization capability of our framework on spoken language understanding task.}
%     \label{tab:abl-2}
% \end{table}

% \subsubsection{Robustness -- Minimal change in final WER with different Objective Function}
% To quantify the variation in model's performance, we 
%     \begin{table}[t]
%         \centering
%         \begin{tabular}{ccccc}
%             \toprule
%             \textbf{$n_\text{DS}$} & \textbf{$RD_{sc}$} & \textbf{$M_{kld+ctc}$} & \textbf{$M_{80-20}$}& \textbf{Ours} \\ \midrule
%             12 & 8.07 & 6.18 & & 5.82 \\
%             10 & 7.28 & 6.29 & & 5.90 \\
%             8 & 7.28 & 6.62 & & 6.32 \\
%             6 & 8.39 & 7.65 & & 7.32 \\
%             4 & 12.81 & 12.79 & & 11.81 \\
%             2 & 39.87 & 40.15 & & 38.30 \\ \bottomrule
%         \end{tabular}
%         \label{tab:3}
%         \caption{WER comparison on Conformer based model for Input Dependent Dropping}
%     \end{table}

\vspace{-1.5mm}
\section{Conclusion}
\vspace{-1mm}
This work introduces an effective framework exploiting KD for curating dynamic speech architectures. Proposed DLD framework harnesses the rich semantic knowledge of teacher network embodied in its latent embeddings to supervise the learning of dynamic student network. The student network learns to dynamically adapt to different encoder sizes in parallel to minimizing the difference between embedding's distribution, hence producing effective latent embeddings. Extensive experimentation using conformer and wavlm architectures underlines the superior performance-computation trade-off as compared to state-of-the-art methods.

\section{Acknowledgments}
This work was supported by Ministero delle Imprese e del Made in Italy (IPCEI Cloud DM 27 giugno 2022 - IPCEI-CL-0000007) and European Union (Next Generation EU).

\bibliographystyle{IEEEbib}
\bibliography{refs}

\end{document}